\documentclass[journal, letterpaper]{IEEEtran}

\usepackage{graphicx}

\usepackage{url}

\usepackage{amsmath}

\usepackage{textgreek}	
\usepackage{listings}
\usepackage{csvsimple}
\usepackage{longtable}

\begin{document}

	\title{Fine-tuning convergence model in Bengali speech recognition}
	\author{Zhu Ruiying, Shen Meng}	
	\maketitle

\begin{abstract}
Research on speech recognition has attracted considerable interest due to the difficult task of segmenting uninterrupted speech. Among various languages, Bengali features distinct rhythmic patterns and tones, making it particularly difficult to recognize and lacking an efficient commercial recognition method. In order to improve the automatic speech recognition model for Bengali, our team have chosen to utilize the wave2vec 2.0 pre-trained model, which has undergone convergence for fine-tuning. Regarding Word Error Rate (WER), the learning rate and dropout parameters were fine-tuned, and after the model training was stable, attempts were made to enlarge the training set ratio, which improved the model's performance. Consequently, there was a notable enhancement in the WER from 0.508 to 0.437 on the test set of the publicly listed official dataset. Afterwards, the training and validation sets were merged, creating a comprehensive dataset that was used as the training set, achieving a remarkable WER of 0.436. 

\end{abstract}

\section{Introduction}
	
Transcription and speech recognition is a typical application of machine learning that has promising commercial prospects. Prominent commercial products for speech recognition and transcription include Siri, Xunfei Hear, and numerous other household-name speech recognition systems. However, these mature commercial products are typically restricted to resource-rich languages of the globe. While speech recognition technology for Bengali, one of the most widely spoken languages, is still in its early stages. Bengali can be classified as a language with limited resources. The competition called \textit{'Bengali.AI Speech Recognition'}\cite{bengaliai-speech}, hosted by Kaggle, aims to enhance speech recognition models specifically for the Bengali language.
 
while managing the problems of speech recognition, the initial inclination is to apply Recurrent Neural Network (RNN). However, RNN has challenges such as vanishing gradients and substantial processing expenses due to the absence of parallelism when handling lengthy data.   Conversely, the Transformer is a model design that does not rely on the structure of a recurrent neural network (RNN). Instead, it leverages general information and does not need word order information, which is crucial in natural language processing (NLP).   It avoids the need of recursion and instead entirely relies on an attentional mechanism to establish connections between inputs and outputs. The Transformer model enables enhanced parallelization and introduces a revolutionary level of sequence-to-sequence conversion. 
    
The most recent development in speech and language technology is the pre-training of extremely large models that can be fine-tuned for particular tasks; therefore, this project proposed wav2vec2.0, which combines the Transformer and the quantization module of vq-wav2vec. \cite{ASR} The encoder network in wav2vec2.0 utilizes a Convolutional Neural Network (CNN). The context network is built upon the Transformer architecture, regarding the aim of reconstructing the quantized frames that have been masked at the feature level. During pre-training on unlabelled speech, the model acquires knowledge of distinct speech units. The model is further refined using the labelled data by employing the connection time classification (CTC) loss. This contributes to subsequent speech recognition tasks. The application of a CNN-based speech embedding method on the wav2vec2.0 model is language-agnostic, as it provides the capability to acquire shared patterns among all spoken languages.

Thus, we select the fine-tuned model \textit{Wave-2-Vec-2-Bengali-Ai} as our pre-trained model. Due to the chosen model having undergone prior fine-tuning, it remains susceptible to issues like overfitting. In turn, additional tuning of the model structure is essential in order to optimize the model's performance and accomplish the aim of WER reduction.

\section{Dataset}
The model training and testing in this project utilize the OOD-Speech dataset\cite{dataset1}. The OOD-Speech dataset is the initial baseline dataset for Automatic Speech Recognition (ASR) in Bengali. It contains 1177.94 hours of training data and 23.03 hours of testing data. The dataset was obtained via online crowdsourcing campaigns, TV shows, and other literary works.  
     
Therefore, this dataset demonstrates a notable variance in distribution between the training and test sets, making it ideal for evaluating the resilience of the ASR model to distributional changes.   During the model training phase, we partitioned a dataset consisting of 706,689 samples. We separated this dataset into separate training and validation sets using different ratios. Furthermore, the kaggle platform offered a test set to evaluate the performance of the model.	
 
\section{Methodology}
\subsection{Pre-trained Models}	
\subsubsection{Model Architecture}
Wave2vec2.0 is a self-supervised learning system designed for speech-audio data. It employs a substantial quantity of unlabelled data to learn representations and then fine-tunes on labelled data. This approach has shown remarkable success in many language recognition tasks. 
 
\begin{figure}[h]
    \centering
    \includegraphics[width=0.5\textwidth]{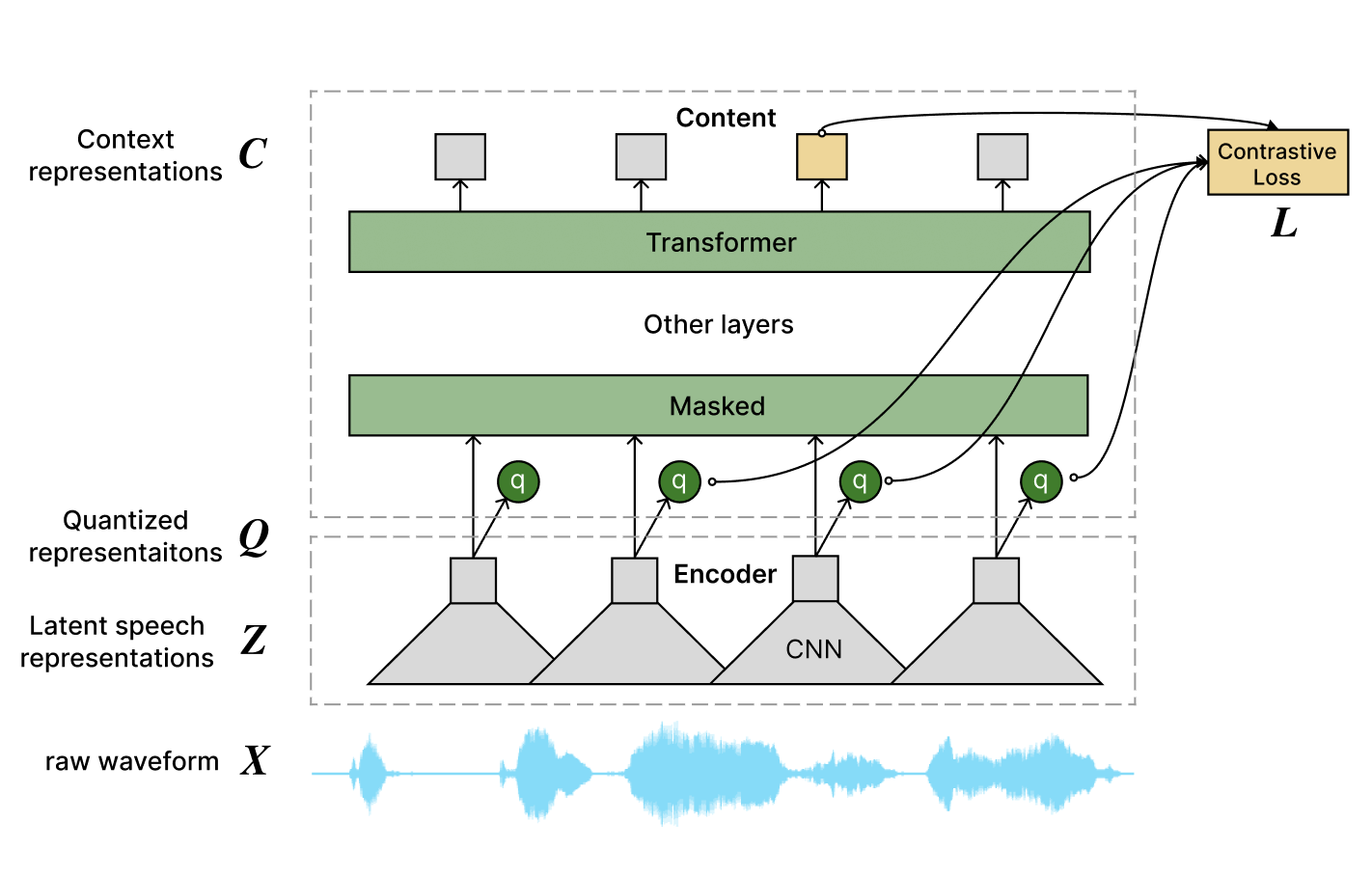}
    \caption{Structure of Wav2vec2.0}
    \label{wav2vec}
\end{figure}

The algorithm's model structure combines a multilayer convolutional feature encoder with a contextualized Transformer network for voice processing applications. \cite{wav2vec2}

Initially, the raw audio is processed through a sequence of convolutional blocks to extract alternative speech representations. The model applies GELU activation and layer normalization methods to enhance the extraction of features. Afterwards, the Transformer network effectively captures the contextual information in the sequence data by utilizing relative position embedding instead of fixed position embedding. Ultimately, the quantization module converts the output of the feature encoder into a limited number of speech representations and uses Gumbel softmax to choose codebook entries in a differentiable manner. This integrated architecture efficiently executes the complete processing and categorization of speech representations, offering robust modeling capabilities for applications connected to speech.

\subsubsection{Model Selection}
Currently, wav2vec2 has an average BER of 1.4\%/2.6\% on the LibriSpeech test/test-other set \cite{zhang2022pushing}. Therefore, it is a natural choice for Bengali ASR.

In this research, we have selected three different Bengali language recognition models that are based on Wave2vec2.0. We then evaluated these models by analyzing their loss.

The three chosen pre-trained models are as follows:
\begin{enumerate}
    \item A model that has undergone fine-tuning and training on other comparable datasets, known as \textit{bengali-ex006}.
    
    \item A fine-tuned model that is based on the competition dataset, denoted as \textit{wave-2-vec-2-bengali-ai}.
    
    \item The self-supervised model \textit{ai4bharat-indicwav2vec-v1-bengali}.

\end{enumerate}

\begin{table}[ht]
\centering
\caption{Model Performance Comparison}
\begin{tabular}{|c|c|c|}
\hline
Model & Training Loss & Validation Loss \\
\hline
bengali-ex006 & 0.9345 & 0.804495 \\
wave-2-vec-2-bengali-ai & 0.7128 & 0.626877 \\
indicwav2vec\_v1\_bengali & 2.3764 & 2.470925 \\
\hline
\end{tabular}
\end{table}

Based on the test comparison depicted in the figure, we observed that training on the converged model \textit{wave-2-vec-2-bengali-ai} significantly decreases the loss when evaluated against the competition's public dataset in the Kaggle competition scoring test. To enhance the model, one can freeze the model layer and then fine-tune it by adjusting the parameters. By fine-tuning the model in succeeding rounds, we may effectively address the issue and improve the migration rate.

\section{Training}
In light of the competition's time limitations and the model's refined performance as determined by the convergent competition dataset, modifications were made to the dataset and validation set distributions throughout the training phase. In order to facilitate the training and validation processes, an approximate 10\% subset of the initial 700,000 data points was sampled from the "valid" subset and appended to the validation set; the remaining 90\% of the data points were allocated to the training set. Approximately 5\% of the data in the "train" subset was sampled for the validation set, while the remaining 92\% was allocated for the training set. The final training set comprises 57,776 samples, whereas the validation set comprises 5,667 samples; both sets are utilized in the training and validation of the models, respectively.

According to the above information, Table \ref{tablerandom} represents the specifics of the partitioning and sampling of the datasets:

\begin{table}[ht]
\centering
\caption{Dataset Partitioning and Sampling}
\begin{tabular}{|c|c|c|}
\hline
Data Subset & Data Count & Allocation Percentage \\
\hline
Training Set & 57776 & 91.4\% \\
Validation Set & 5667 & 8.6\% \\
\hline
\end{tabular}
\label{tablerandom}
\end{table}

\section{Fine-tuning}
During the analysis of the outcomes, it was discovered that there was excessive convergence, leading to challenges in enhancing the test findings.   To address the issue of overfitting, the following approaches were employed to optimize the training model. 

\subsubsection{Dropout}
The research utilizes data with notable out-of-distribution attributes, and appropriately modifying the dropout probability can mitigate the overfitting of the model and enhance its resilience to distribution shifts.

\subsubsection{Learning rate}
By implementing a progressive learning rate adjustment approach, in the early stages of the training period, when the model's performance remains unreliable, a more rapid convergence to a point near the minima is achieved through the use of a higher learning rate. Reduce the learning rate progressively as the model's loss tends to stabilize and decelerate, enabling the model to acquire more intricate features.

\subsubsection{Dataset scaling}
After achieving stability in model training, we aim to increase the fraction of the training set and retrain the model using the same parameters. This enables the model to extract information from a larger amount of data, hence enhancing its performance.

\section{Results}

\textbf{The learning rate (lr)}: According to Table \ref{table1}, it is clear that when the lr is set to 1.00E-05, independent of the dropout parameters, the WER remains relatively constant at around 0.438. Conversely, in Table \ref{tabler2}, when the value of lr is modified within the range of 2.00E-05 to 1.00E-07, the WER constantly stays at 0.439. These data indicate that the performance of the model remains unaffected by fluctuations in the learning rate, even when it is altered throughout a wide range, given the current hyperparameter settings.

\textbf{The Dropout parameter}: In Table \ref{table1}, when the learning rate (lr) is set to 1.00E-05, variations in the attention dropout and hidden layer dropout have a minimal impact on the WER, changing only within the range of 0.437 to 0.441. This suggests that the model is resistant to the dropout parameter, and variations within a specific range have a relatively minimal effect on the model.

\begin{table}[ht]
\centering
\caption{Model Comparison with Different Dropout Parameters}
\begin{tabular}{|c|c|c|c|}
\hline
lr & Attention Dropout & Hidden Dropout & WER \\
\hline
1.00E-05 & 0.1 & 0.05 & 0.441 \\
1.00E-05 & 0.1 & 0.1 & 0.439 \\
1.00E-05 & 0.1 & 0.15 & 0.438 \\
1.00E-05 & 0.1 & 0.2 & 0.438 \\
1.00E-05 & 0.2 & 0.1 & 0.437 \\
1.00E-05 & 0.2 & 0.15 & 0.438 \\
1.00E-05 & 0.2 & 0.2 & 0.437 \\
1.00E-05 & 0.2 & 0.2 & 0.436 \\
\hline
\end{tabular}
\label{table1}
\end{table}

\begin{table}[ht]
\centering
\caption{Model Comparison with Different Learning Rates}
\begin{tabular}{|c|c|c|c|}
\hline
lr & Attention Dropout & Hidden Dropout & WER \\
\hline
2.00E-05 & 0.1 & 0.1 & 0.439 \\
1.00E-05 & 0.1 & 0.1 & 0.439 \\
1.00E-06 & 0.1 & 0.1 & 0.439 \\
1.00E-07 & 0.1 & 0.1 & 0.439 \\
\hline
\end{tabular}
\label{tabler2}
\end{table}

Collectively, the existing statistics suggest that modifications in the learning rate and dropout parameter have a rather restricted influence on the model's performance within the current configuration. It might be essential to broaden the search space for hyperparameters and generate additional hyperparameters for fine-tuning in order to discover a more effective combination of hyperparameters. Furthermore, in order to enhance the stability and resilience of the model, it is important to conduct experiments and make adjustments on a bigger scale to achieve superior model performance.

\section{Conclusion}
In this competition, we achieved a marginal improvement of 0.05, reducing our score from 0.441 to 0.436. As a result, we secured the 95th position on the public leaderboard and the 113th position on the private leaderboard of the Kaggle competition. 

To enhance the automatic voice recognition model for Bengali, we have opted to refine the wave2vec 2.0 pre-trained model specifically for Bengali language recognition. In order to tackle the diversity of this language and the uneven distribution of data, we have implemented a range of efficient optimization strategies.

To tackle the stability issue during model training, we employ a progressive learning rate adjustment strategy. This involves gradually modifying the learning rate to ensure rapid convergence of the model in the initial phase of training. Subsequently, the feature learning process is gradually refined to enhance recognition accuracy. Furthermore, when taking into account the out-of-distribution characteristics of the data, we systematically raise the dropout probability. This approach substantially reduces the occurrence of overfitting and simultaneously enhances the model's ability to handle variations in data distribution. Subsequently, once the model training reached a stable state, we increased the ratio of the training set in the dataset. This was done to enable the model to learn from a larger amount of data, resulting in significant enhancements to its performance.

The model was trained and tested on OOD-Speech, a benchmark dataset for Bengali ASR that contains out-of-distribution samples. It achieved a remarkable Word Error Rate (WER) score of 0.436 on the test dataset provided by Kaggle. This result showcases the model's exceptional ability to handle the diverse Bengali dialects and the inconsistent data distribution, highlighting its robustness and accuracy. Our work in Bengali ASR modeling fills a significant void in the field, offering crucial assistance and direction for the advancement and implementation of Bengali ASR techniques. Our research and practice not only establish a strong basis for the future advancement of Bengali ASR modeling but also offer valuable guidance and expertise for addressing analogous challenges in other languages.

\section{Discussion}

By assimilating the mindset of the medalist, our team intends to use the following strategies: 

\begin{enumerate}

    \item Combine artificial intelligence via the OpenAI Whisper-medium model.   Whisper is a pre-trained model based on the Transformer architecture that is specifically designed for tasks involving Automatic voice Recognition (ASR) and voice translation.   The model is trained on a dataset of 680,000 hours of labeled speech using a large-scale weakly supervised approach. It shows significant generalization capabilities and can be used to various datasets and domains without the need for fine-tuning.   This model has superior performance in the challenge of Bengali recognition. 
    
    \item Expand the datasets to include Datasets: OpenSLR 37, OpenSLR 53, MadASR, Shrutilipi, Macro, Kathbath, GoogleTTS generated audios, and pseudo-labeled YouTube videos. However, ensure that they are also checked for training length and mobility. 
    
    \item Apply the Punctuation Prediction Model \textit{ai4bharat/IndicBERTv2-MLM-Sam-TLM}. After training with LSTM head and data enhancement techniques, it becomes a powerful model that successfully predicts particular punctuation marks in Bengali. To enhance the overall performance and resilience of the model, it is advisable to integrate three distinct subsets of the training model and utilize the beam search decoding approach during the inference phase. The approach has demonstrated remarkable efficacy.

\end{enumerate}

\bibliographystyle{unsrt} 
\bibliography{reference_kaggle}  

\end{document}